\documentclass[12pt,preprint,notoc,nohyper]{ibrarp4} 

\usepackage{epsfig,multicol,bbm}

\newcommand\fverb{\setbox\pippobox=\hbox\bgroup\verb}
\newcommand\fverbdo{\egroup\medskip\noindent%
            \fbox{\unhbox\pippobox}\ }
\newcommand\fverbit{\egroup\item[\fbox{\unhbox\pippobox}]}
\newbox\pippobox

\title{Weyl collineations that are not curvature collineations}

\author{Ibrar Hussain$^a$, Asghar Qadir$^a$ and
K. Saifullah$^b$ \\

$^a$Centre for Advanced Mathematics and Physics,
National University of Sciences and Technology, Rawalpindi, Pakistan  \\

$^b$Department of Mathematics, Quaid-i-Azam University,
Islamabad, Pakistan \\

Electronic address: \email{ibrar\_msw@yahoo.com},
\email{aqadirs@comsats.net.pk}, \email{saifullah@qau.edu.pk}}

\preprint{}  

 \abstract{Though the Weyl tensor is a
linear combination of the curvature tensor, Ricci tensor and Ricci
scalar, it does not have all and only the Lie symmetries of these
tensors since it is possible, in principle, that ``asymmetries
cancel''. Here we investigate if, when and how the symmetries can be
different. It is found that we can obtain a metric with a finite
dimensional Lie algebra of Weyl symmetries that properly contains
the Lie algebra of curvature symmetries. There is no example found
for the converse requirement. It is speculated that there may be a
fundamental reason for this lack of ``duality''.}

\dedicated{}

\begin{document}

\section{Introduction}

In general relativity the Ricci tensor and Ricci scalar combine to
give the matter content of the spacetime and the Weyl tensor gives
the gravitational field with the matter content removed \cite{PR}.
As such the Weyl tensor plays a fundamental role in understanding
the purely gravitational field for a given metric. Since it is
conformally invariant \cite{lss}, i.e. remains unchanged under
infinitesimal re-scalings, its local symmetries are of particular
interest. Local symmetries of the metric tensor, called \textit{%
isometries} or \textit{Killing vectors} (KVs), are given by
\begin{equation}\label{1}
\pounds _{_{\mathbf{X}}}\mathbf{g}=0,
\end{equation}
where $\pounds _{_{\mathbf{X}}}$ is the Lie derivative along the
vector field $\mathbf{X}$ and $\mathbf{g}$ is the metric tensor.
Replacing $\mathbf{g}$ by any tensor field gives the local
symmetries of that tensor, called \textit{collineations} \cite{kit}.
Putting $\lambda \mathbf{g}$ on the right side of Eq. (\ref{1}),
gives the \textit{conformal Killing vectors} (CKVs). If $\lambda $
reduces from being any (differentiable) function to a constant
number, $\mathbf{X}$ is called a \textit{homothetic vector} (HV) or
a \textit{homothety}. The proper solution of the non-homogeneous
equation is called a \textit{proper} HV, while the complementary
function gives linear combination of KVs. The complete general
solution gives the set of HVs, \{HVs\}, which contains \{KVs\}
properly if there exists a proper HV and otherwise \{HVs\}$\equiv
$\{KVs\}. Special significance attaches to \{HVs\} as they are the
Noether symmetries \cite{stp} of the Einstein-Hilbert Lagrangian
$\sqrt{\left| g\right| }R,$ where $R$ is the Ricci scalar \cite
{lsl}. Since the metric tensor is everywhere non-singular, \{KVs\}
form a finite dimensional Lie algebra of dimension $\leq n(n+1)/2$
for a manifold of dimension $n.$

The curvature tensor or Ricci tensor can be ``degenerate'' in the
sense that their ``determinant'' is zero. (The 4th rank curvature
tensor of 4-dimensions can be represented by a 6 dimensional matrix,
on account of its algebraic symmetries, whose rank gives the rank of
the tensor.) In this case the system of equations can become
under-determined and the resulting Lie algebra can become infinite
dimensional. By Noether's theorem \cite{RBM} the
\textit{homotheties} give the conservation laws for the spacetime
given by $\mathbf{g.}$ Clearly, special interest attaches to the
case where there are finitely many conserved quantities and hence
the Lie algebra of \textit{curvature
collineations} \{CCs\} is finite dimensional. Clearly \{CCs\}$\supseteq $%
\{HVs\}.

The Weyl tensor, $\mathbf{C}$, can be written in components form as
\[
C_{cd}^{ab}=R_{cd}^{ab}-2\delta _{\lbrack
c}^{[a}R_{d]}^{b]}+\frac{1}{3}\delta _{\lbrack c}^{a}\delta
_{d]}^{b}R,
\]
where $R_{bcd}^{a}$ is the curvature tensor $R_{ab}$ is the Ricci
tensor. It is trace-free. Replacing $\mathbf{g}$ by $\mathbf{C,}$ in
component form Eq. (\ref{1}) becomes
\[
C_{bcd,f}^{a}X^{f}+C_{fcd}^{a}X_{,b}^{f}+C_{bfd}^{a}X_{,c}^{f}+C_{bcf}^{a}X_{,d}^{f}-C_{bcd}^{f}X_{,f}^{a}=0,
\]
where ``$,$'' denotes the partial derivative. Though $\mathbf{C}$
and the curvature tensor have similar forms, the local symmetries of
the curvature tensor (i.e. CCs) and \textit{Weyl collineations}
(WCs) are different. There has been very little work done which even
mentions WCs \cite{kit,gs1,gs2,ahb,shb}, some of which has errors as
mentioned in the conclusion. Indeed if the Ricci tensor,
$\mathbf{R,}$ is zero, i.e. for vacuum with zero cosmological term,
\{WCs\} $\equiv $ \{CCs\}, as the Weyl tensor reduces to the
curvature tensor \cite{kit}. The tensor $\mathbf{R}$ is degenerate\
if its matrix in any coordinate basis is of rank $3$ or less.
Clearly, it is possible to arrange that one be degenerate without
the other being degenerate. For example, if the spacetime is of
Petrov type O the Weyl tensor is zero (except for Minkowski space)
while the curvature tensor is not. We can choose a metric of type O
with non-degenerate curvature tensor. Then the Lie algebra of
\{CCs\} will be of dimension less then or equal to 6 and of \{WCs\}
infinite dimensional such that every vector field is a WC. An
example is the De-Sitter (or anti De-Sitter) spacetime.

The question arises whether the case of a finite dimensional Lie
algebra of \{WCs\} and an infinite dimensional Lie algebra of
\{CCs\}, can be found. It is not a priori obvious that it will,
since the curvature tensor has up to 10 independent components while
the Weyl tensor has only 6 (due to the trace-free condition). In
this paper we have investigated the relation between the \{WCs\} and
\{CCs\} in specific cases, with a view to finding more general
statements about the relation. Where examples are found the
existence of such metrics is obviously proved but when they are not
found it does not prove that they do not exist. Better methods would
be needed to obtain the final answer in that case. It would be of
interest to obtain answers to these questions for at least some
classes of metrics.

\section{Examples of unequal \{WCs\} and \{CCs\}}

The simplest attempt to find the desired examples, is to consider a
non-vacuum spacetime. The examples that spring to mind are the
Schwarzschild interior and Reissner-Nordstrom metrics \cite{mtw}. In
the former the Ricci scalar is non-zero while in the latter the
Ricci scalar is zero but the Ricci tensor is non-zero. The
Schwarzschild interior solution is Petrov type O \cite{ese} and thus
every vector field is a WC while KVs, HVs and CCs are four. The
Reissner-Nordstrom spacetime is of Petrov type D and there are the
same four WCs, KVs, HVs and CCs. But when we take pressure as
constant the Schwarzschild interior has the following non-zero
components of curvature and Weyl tensor
\[R_{212}^{1}=\frac{8\pi Gp}{c^{4}}r^{2}, R_{313}^{1}=R_{212}^{1}\sin
^{2}\theta\]

\begin{eqnarray*}
C_{101}^{0} &=&\frac{k(3r-1)}{r(1-kr^{2})}, C_{202}^{0}=-\frac{1}{6}%
kr(3r+5), C_{212}^{1}=\frac{1}{2}kr(3r-\frac{5}{3}), \\
C_{303}^{0} &=&C_{202}^{0}\sin ^{2}\theta ,C_{313}^{1}=C_{212}^{1}%
\sin ^{2}\theta , C_{323}^{2}=-\frac{4k}{3}r\sin ^{2}\theta ,
\end{eqnarray*}
where $k=8\pi Gp/c^{4}$. In this case the CCs are arbitrary and
\{WCs\} = \{KVs\} = \{HVs\} = 4, with generators given by
\begin{eqnarray*}
\mathbf{X}_{0} &=&\frac{\partial }{\partial t}, \mathbf{X}_{1}=-\sin
\phi \frac{\partial }{\partial \theta }-\cot \theta \cos \phi
\frac{\partial
}{\partial \phi }, \\
\mathbf{X}_{2} &=&\cos \phi \frac{\partial }{\partial \theta }-\cot
\theta
\sin \phi \frac{\partial }{\partial \phi }, \mathbf{X}_{3}=\frac{%
\partial }{\partial \phi }.
\end{eqnarray*}
Thus \{WCs\} is properly contained in \{CCs\}. In the case of
Reissner-Nordstrom metric the Ricci scalar is zero. However, \{KVs\}
= \{HVs\} = \{CCs\} = \{WCs\} given by the Lie group
$G_{4}=SO(3)\otimes \Bbb{R}$ (where $\otimes $ denotes direct
product) with the generators given above.

Looking through the complete classification of spherically symmetric
static metrics by KVs, CCs and RCs \cite{aqr-mz,ahb-ark-aq,mz} did
not yield any interesting case. We, therefore, looked at the
corresponding classification of cylindrically symmetric static
spacetimes \cite{Aq-Mz,Ahb-Ark-Aq,aq-ks-mz} and plane symmetric
static spacetimes \cite{Tf-Aq-Mz,AHB-ARK-AQ,Tbf-aq-mz} for this
purpose.

The general cylindrically symmetric static metric is
\begin{equation}\label{2}
ds^{2}=e^{\nu (r)}dt^{2}-dr^{2}-a^{2}e^{\lambda (r)}d\theta
^{2}-e^{\mu (r)}dz^{2}.
\end{equation}
This metric has 3 KVs in general, which generate the Lie group
$SO(2)\otimes \Bbb{R}\otimes \Bbb{R}$,
\[
\mathbf{X}_{0}=\frac{\partial }{\partial t}, \mathbf{X}_{1}=\frac{1}{a%
}\frac{\partial }{\partial \theta }, \mathbf{X}_{2}=\frac{\partial
}{\partial z},
\]
where, in Eq. (\ref{2}) $a$ is a constant with dimensions of length and $\nu ,$ $%
\lambda $ and $\mu $ are arbitrary functions \cite{ese}. In the case $%
\lambda =$ constant, we get a cylindrical analogue of the
Bertotti-Robinson metrics and we can choose $\lambda =0.$ If this is
not the case we naturally choose the function so that $a$ gets
replaced by $r$ and we are left with some other general function of
$r$. The plane symmetric general metric can be written as \cite{ese}
\begin{equation}\label{3}
ds^{2}=e^{\nu (x)}dt^{2}-dx^{2}-e^{\mu (x)}(dy^{2}+dz^{2}).
\end{equation}
This metric has 4 KVs in general, which generate the Lie group [$%
SO(2)\otimes _{s}\Bbb{R}^{2}]\otimes \Bbb{R}$, (where $\otimes _{s}$
denotes semi-direct product) given by
\[
\mathbf{X}_{0}=\frac{\partial }{\partial t}, \mathbf{X}_{1}=\frac{%
\partial }{\partial y}, \mathbf{X}_{2}=\frac{\partial }{\partial z}, \mathbf{X}_{3}=-z\frac{\partial }{\partial y}+y\frac{\partial }{%
\partial z}.
\]
The first case of interest is when $\nu =0$ in Eq. (\ref{2}) and
$e^{\lambda }=e^{\mu }=(r/a)^{2}$. In this case the stress-energy
tensor is given by
\[
T_{00}=-\frac{1}{\kappa r^{2}}=-T_{11}, T_{22}=0=T_{33},
\]
so that it is not a realistic spacetime. The non-zero component of
the curvature tensor is
\[
R_{323}^{2}=-\frac{1}{a^{2}}
\]
and those of the Weyl tensor are
\begin{eqnarray*}
C_{101}^{0} &=&-\frac{1}{3r^{2}},
C_{202}^{0}=\frac{1}{6}=C_{212}^{1},
\\
C_{303}^{0} &=&\frac{1}{6a^{2}}=C_{313}^{1},
C_{323}^{2}=-\frac{1}{3a^{2}}.
\end{eqnarray*}
Here the Ricci tensor is of rank 2 and is degenerate.

This case has one extra KV
\[\mathbf{X}_{3}=-\frac{z}{a}\frac{\partial }{\partial \theta }+a\theta \frac{\partial }{\partial z},\]
one proper HV
\[
\mathbf{X}_{4}=t\frac{\partial }{\partial t}+r\frac{\partial
}{\partial r},
\]
and one additional WC
\[
\mathbf{X}_{5}=\frac{1}{2}(t^{2}+r^{2})\frac{\partial }{\partial t}+tr\frac{%
\partial }{\partial r}.
\]
There are infinitely many CCs and RCs. Clearly, here \{KVs\}
$\varsubsetneq $ \{HVs\} $\varsubsetneq $\ \{WCs\} $\varsubsetneq $
\{CCs\}.

A case with the reverse inclusion appears for the plane symmetric
metric, with $e^{\nu }=e^{\mu }=(x/a)^{b}$ ($a,b\neq 0\in \Bbb{R}$)
in Eq. (\ref{3}). Here
\[
T_{00}=\frac{bx^{b}}{\kappa 4x^{2}a^{b}}(4-3b)=-T_{22}=-T_{33},
T_{11}=\frac{3b^{2}}{\kappa 4x^{2}},
\]
which represents a realistic spacetime for $0<b<\frac{1}{4}$ and
satisfies the positive energy condition ($T>0$) for
$0<b<\frac{1}{4}.$\ It is clearly an anisotropic spacetime. This is
a Petrov type O metric with non-zero curvature tensor components
\begin{eqnarray*}
R_{101}^{0} &=&\frac{b}{4}(\frac{2-b}{x^{2}}),
R_{212}^{1}=R_{313}^{1}=(\frac{x}{a})^{b}(\frac{b(2-b)}{4x^{2}}), \\
R_{202}^{0}
&=&\frac{-b^{2}x^{b}}{4a^{b}x^{2}}=R_{303}^{0}=R_{323}^{2}.
\end{eqnarray*}
This spacetime has two extra KVs
\[
\mathbf{X}_{4}=z\frac{\partial }{\partial t}+t\frac{\partial }{\partial z},
\mathbf{X}_{5}=y\frac{\partial }{\partial t}+t\frac{\partial }{%
\partial y},
\]
and one proper HV which is also a CC and RC
\[
\mathbf{X}_{6}=(t\frac{\partial }{\partial t}+y\frac{\partial }{\partial y}+z%
\frac{\partial }{\partial z}), (b\neq 2).
\]
In case $b=2$, the curvature tensor becomes degenerate and the Lie
algebra of \{CCs\} infinite dimensional. However, not every vector
field will be a CC, while every vector field is a WC. Clearly
\{CCs\} $\varsubsetneq $ \{WCs\} for this metric (for all $b$).

A more interesting case is $e^{\nu }=(r/a)^{4},$ $e^{\lambda
}=e^{\mu }=(r/a)^{2}$ in Eq. (\ref{2}). Here
\[
T_{00}=-\frac{r^{2}}{\kappa a^{4}}, T_{11}=\frac{5}{\kappa r^{2}},
T_{22}=\frac{4}{\kappa }, T_{33}=\frac{4}{\kappa a^{2}},
\]
which is also a non-realistic spacetime. The non-zero components of
the curvature tensor are
\[
R_{101}^{0}=-\frac{2}{r^{2}}, R_{202}^{0}=2,
R_{303}^{0}=\frac{-2}{a^{2}},R_{323}^{2}=-\frac{1}{a^{2}},
\]
and those of the Weyl tensor are
\begin{eqnarray*}
C_{101}^{0} &=&\frac{1}{3r^{2}},
C_{202}^{0}=\frac{1}{6}=C_{212}^{1},
\\
C_{303}^{0} &=&\frac{1}{6a^{2}}=C_{313}^{1}=-C_{323}^{2}.
\end{eqnarray*}
This case has one extra KV
\[
\mathbf{X}_{3}=-\frac{z}{a}\frac{\partial }{\partial \theta
}+a\theta \frac{\partial }{\partial z},
\]
one proper HV which is also a CC
\[
\mathbf{X}_{4}=-t\frac{\partial }{\partial t}+r\frac{\partial
}{\partial r},
\]
and one additional WC which is also a conformal vector field
\[
\mathbf{X}_{5}=-\frac{1}{2}(\frac{b^{4}}{r^{2}}+t^{2})\frac{\partial }{%
\partial t}+tr\frac{\partial }{\partial r}.
\]
Here $\mathbf{X}_{3}$ gives a local spatial rotation between the
axial and rotational symmetry directions and $\mathbf{X}_{4}$ and
$\mathbf{X}_{5}$ are
scaling symmetries. Clearly $\left\langle \mathbf{X}_{1},\mathbf{X}_{2},%
\mathbf{X}_{3}\right\rangle $ gives the plane symmetry group and $[\mathbf{X}%
_{0},\mathbf{X}_{4}]=-\mathbf{X}_{0},$ $\mathbf{[\mathbf{X}}_{0}\mathbf{,%
\mathbf{X}}_{5}\mathbf{]=-X}_{4},$ $[\mathbf{X}_{4},\mathbf{X}_{5}]=-\mathbf{%
X}_{5},$ while \{CCs\} $\varsubsetneq $\ \{WCs\} $\equiv$\ \{RCs\}.

\section{Conclusion}

We found only two papers that address Weyl collineations properly,
both of which have errors and address only limited aspects of the
problem \cite {ahb,shb}. For example in \cite{ahb} it is claimed
that there are 10 WCs for De-Sitter and anti De-Sitter metrics. This
is obviously wrong as these spaces are of Petrov type O, for which
all vector fields are WCs. Again in Ref. 10 only infinite
dimensional Lie algebras for Weyl collineations of pp-waves are
found. In that paper $R_{11}$ is taken to be the only non-zero
component of the Ricci tensor, while for pp-waves $R_{00}\neq 0.$
Correcting the error does not provide any interesting example.

In the present paper we have found non-trivial examples of WCs that
are not simply CCs. Of course, the Petrov type O example is trivial
in another sense, namely that all vector fields are Weyl
collineations, thus \{CCs\} $\varsubsetneq $ \{WCs\}. The first
example of cylindrical symmetry discussed in this paper was trivial
in yet another sense, namely that the Lie algebra of curvature
collineations is infinite dimensional. Thus \{WCs\} $\varsubsetneq $
\{CCs\}. However the last case is entirely non-trivial as it has
\{CCs\} $\varsubsetneq $ \{WCs\} and both have finite dimensional
Lie algebras. The question arises whether there are cases in which
\{WCs\} $\varsubsetneq $ \{CCs\} and both have finite dimensional
Lie algebras.

It is possible that the rank of the $6 \times 6$ Weyl matrix is
greater or less than the rank of the corresponding curvature matrix.
If the rank of the curvature matrix is $\geq 4$  then the Lie
algebra of CCs is finite dimensional \cite{Ahb-Ark-Aq}. From here it
seems possible that there do exist cases with \{WCs\}
$\varsubsetneq$ \{CCs\} that remain finite dimensional. However, the
process of calculation seemed to indicate that such cases may not be
possible. It would be worth while to either find such a case or to
provide\ a definite proof that it does not exist. If it does not,
the ``duality'' between curvature and Weyl collineations that may
have been expected would be shown to be violated.

\acknowledgments

Useful discussions with Ugur Camci are acknowledged. IH would like
to thank the Higher Education Commission of Pakistan and
Quaid-i-Azam University, Islamabad for financial support provided
during this work.

\end{document}